\documentclass[a4paper,twocolumn,aps,prl,nolongbibliography,superscriptaddress,amsmath,amssymb,floatfix,nofootinbib]{revtex4}

\usepackage{graphicx}
\usepackage{dcolumn}
\usepackage{bm}
\usepackage{amsmath}
\usepackage{epstopdf}
\usepackage{amsfonts}
\usepackage{amssymb}
\usepackage{epsfig}
\usepackage{tabularx}
\usepackage{color}

\usepackage[colorlinks=true,citecolor=blue,urlcolor=magenta,breaklinks]{hyperref}

\newcommand{\be}{\begin{equation}}
\newcommand{\en}{\end{equation}}
\newcommand{\bea}{\begin{eqnarray}}
\newcommand{\ena}{\end{eqnarray}}

\begin{document}

\title{White dwarf cooling via gravity portals}
	
\author{Grigoris Panotopoulos} 
\email{grigorios.panotopoulos@tecnico.ulisboa.pt}
\affiliation{Center for Astrophysics and Gravitation, Department of Physics,\\ Instituto Superior T\'ecnico, University of Lisbon, Av. Rovisco Pais 1, 1049-001 Lisboa, Portugal}
	
\author{Il{\'i}dio Lopes} 
\email{ilidio.lopes@tecnico.ulisboa.pt}
\affiliation{Center for Astrophysics and Gravitation, Department of Physics,\\ Instituto Superior T\'ecnico, University of Lisbon, Av. Rovisco Pais 1, 1049-001 Lisboa, Portugal}

\begin{abstract}
We investigate the impact of gravity portal to properties of white dwarfs, such as equation-of-state as well as cooling time. We find that the interaction between dark matter spin-zero bosons and electrons in the mean field approximation softens the equation-of-state, and the object coolers down slower compared to the usual case.
\end{abstract}

\maketitle

%%%%%%%%%%%%%%%%%%%%%%%
\section{Introduction}
%%%%%%%%%%%%%%%%%%%%%%

Multiple current observational data coming from Cosmology and Astrophysics indicate that
ordinary, luminous matter comprise only a tiny percentage of the total energy budget of the Universe \cite{planck}. Einstein's General Relativity (GR) \cite{einstein} can be made compatible with modern data only if another form of matter, which does not have neither strong nor electromagnetic interactions, and manifests itself via gravity, is postulated to exist. This new form, now called dark matter (DM), is non-relativistic in nature, roughly 5 times more abundant than baryonic matter \cite{planck}, and it should be searched for in extension of the Standard Model (SM) of Particle Physics. Its nature and origin still remains a mystery, and it comprises one of the biggest challenges if modern theoretical Cosmology and Particle Physics. For reviews see e.g. \cite{DM1,DM2,DM3,DM4,DM5} and references therein, and for a list of good DM candidates see \cite{taoso}.

\smallskip

Since the spin of the particle that plays the role of DM in the Universe is still unknown, the simplest way to extend the SM to include a good DM candidate is to introduce a scalar field, i.e. a spin-zero boson. Scalar fields are simpler in their treatment as they do not carry neither spinor nor Lorentz indices, and they arise in many different set ups in modern Particle Physics, for instance i) the Higgs sector needed to break electroweak symmetry and to give masses to particles \cite{higgs1,higgs2}, ii) pseudo-Goldstone bosons (pNGB) associated with explicit breaking of additional global symmetries \cite{freese}, iii) moduli from superstring theory compactifications \cite{moduli1,moduli2,moduli3,moduli4,moduli5}, iv) supermultiplets in supersymmetric theories \cite{martin} and supergravity \cite{nilles} contain several scalar fields etc, to mention just a few. 

\smallskip

There is one more reason why one may consider the possibility of DM consisting of a scalar field. The $\Lambda$CDM model, based on cold DM and a positive cosmological constant, has become the concordance cosmological model. Dark matter in the standard parametrization is assumed to be made of weakly interacting massive particles, a conjecture which works very well at large (cosmological) scales ($\ge Mpc$), but unfortunately at smaller (galactic) scales a few problems arise, such as the missing satellites problem, the core-cusp problem, and the too-big-to-fail problem \cite{2017arXiv170502358T}. These problems may be tackled in the context of self-interacting dark matter \cite{spergel1,spergel2}, as any cuspy feature will be smoothed out by the 
dark matter collisions. In addition, if dark matter consists of ultralight scalar particles with a mass $m \leq eV$, and with a small repulsive quartic self-interaction a Bose-Einstein condensate (BEC) may be formed with a long range correlation. This scenario has been proposed as a possible solution to the aforementioned problems at galactic scales \cite{proposal1,proposal2,proposal3}. For a review see e.g. \cite{review}.

\smallskip

Since the presence of DM is inferred only via gravitational interactions, it seems more than natural to couple the Lagrangian of Particle Physics to GR. Then one possibility that should not be ignored is a nonminimal coupling of the scalar field to gravity. Higgs inflation is a notable example of this type of scenario \cite{shapo1,shapo2}. After all, as is well-known since long time ago, even if this term is absent at tree level it will be generated via quantum loop corrections \cite{faraoni}. Even if the scalar field that plays the role of DM does not have any direct interactions with the SM fields in the Jordan frame, its nonminimal coupling to gravity will induce non-vanishing interaction vertices in the Einstein frame after a conformal transformation is performed, see the discussion in the next section. The predictions and the consequences of the gravity portal in the lifetime of the DM particle for different values of its mass and its nonminimal coupling have been investigate in \cite{portal1,portal2}.

\smallskip 

In the present work we propose to investigate for the first time the impact of the gravity portal on properties of white dwarfs (WD), such as equation-of-state (EoS) and cooling time. Being compact enough, WDs serve as ideal stellar laboratories for new gravitational effects. Their advantage over other compact objects, e.g. neutron stars, is that their equation-of-state is relatively well-understood. The Fermi pressure of the degenerate electron gas prevents the collapse of the star due to its own gravity, and thus hydrostatic equilibrium is achieved. For these reasons the choice of WDs as cosmic laboratories for gravity is quite popular in the literature, see e.g. \cite{kouvaris,saltas}.

\smallskip

The plan of our work is the following. In the two subsections of the next section we briefly review the gravity portal scenario, and we summarize the standard EoS of an ideal Fermi gas and the time dependence of the luminosity of WDs. In section 3 we discuss the impact of the gravity portal on the EoS and the cooling time of WDs, and finally we conclude our work in the last section.

%%%%%%%%%%%%%%%%%%%%%%%
\section{Formalism}
%%%%%%%%%%%%%%%%%%%%%%%

%%%%%%%%%%%%%%%%%%%%%%%%%%%%%%%%%%%%%%
\subsection{The gravity portal scenario}
%%%%%%%%%%%%%%%%%%%%%%%%%%%%%%%%%%%%%%

First let us briefly review the gravity portal following \cite{portal1,portal2}.
We extend the Lagrangian of the SM to include dark matter by adding a scalar 
field $\phi$, $\mathcal{L}_T = \mathcal{L}_{SM} + \mathcal{L}_{DM}$, where $\mathcal{L}_{SM}$ is the usual Lagrangian of the SM \cite{SM1,SM2}, while the scalar field is described by the Lagrangian
\begin{equation}
\mathcal{L}_{DM} = \frac{1}{2} \partial_\mu \phi \partial^\mu \phi - V(\phi)
\end{equation}
where the scalar potential includes a mass term and possible self-interactions. For instance, for repulsive dark matter it may have the form \cite{Fan,LP}
\begin{equation} \label{firstkind}
V(\phi) = \frac{1}{2} m_\phi^2 \phi^2 + \frac{1}{24} \: \lambda \phi^4  + ...
\end{equation}
where $m_\phi$ is the mass of the DM particle, and $\lambda = (m_\phi/F)^2 > 0$, with $F$ being a high mass scale, is the self-interaction coupling constant, or for pNGB it has the form \cite{barbieri}
\begin{equation}
V(\phi) = \Lambda^4 [ 1 \pm cos(\phi/F) ]
\end{equation}
which upon expansion around the minimum leads to an attractive force \cite{chavanis}.

In \cite{harko1,harko2} it has been shown that a spin-zero particle can form a Bose-Einstein condensate solving the core/cusp problem at galactic scales.
This, however, requires a repulsive force, and this is why in the following we shall focus on potentials of the first kind, eq. (\ref{firstkind}). Moreover, if it is assumed that the scattering length is of the order of $1~fm$, the mass of the DM particle is computed to be $10~meV$.

Then we couple the particle physics Lagrangian to gravity. In the physical (Jordan) frame the model is described by the action
\begin{equation}
S_J = \int \mathrm{d} ^4x \sqrt{-g} \left[ - \frac{1}{2 \kappa^2} R + \mathcal{L}_T -\xi R f(\phi) \right]
\end{equation}
where $R$ is the Ricci scalar, $g$ the determinant of the metric tensor $g_{\mu \nu}$, $\kappa^2=8 \pi G$, and we allow for a non-minimal coupling to gravity. We remind the reader that this type of coupling is not optional, rather it is inevitable, since it will be generated via quantum loop corrections even if it is absent in the classical action. The functional form of the factor $f(\phi)$ depends on the concrete model considered each time. 

\smallskip

Performing a conformal transformation 
\begin{equation}
\tilde{g}_{\mu \nu} = \Omega^2 g_{\mu \nu}
\end{equation}
where $\Omega^2 = 1 + 2 \xi \kappa^2 f(\phi)$
the action in the Einstein frame takes the form \cite{portal1,portal2}
\begin{equation}
S_E = \int \mathrm{d} ^4x \sqrt{-\tilde{g}} \left[ - \frac{1}{2 \kappa} \tilde{R} + \mathcal{L}_{\phi,SM} +... \right]
\end{equation}
where the dots denote terms that are not of interest here, while the interaction Lagrangian between the SM particles and the dark matter particle $\phi$ is found to be \cite{portal1,portal2}
\begin{equation}
\mathcal{L}_{\phi,SM} = - 2 \xi \kappa^2 \frac{\partial f}{\partial \phi}|_{\phi=0} \phi \left[ \frac{3}{2}\: T_f + ... \right]
\end{equation}
where for the time being we are interested only in the coupling between the spin-zero boson and the SM fermions. 

\smallskip

We see that even if in the Jordan frame there are no direct couplings between $\phi$ and the SM particles, in the Einstein frame there are interaction terms induced due to the nonminimal coupling $\xi$. For heavy DM particles, $m_\phi \geq 1~GeV$, in the gravity portal studied in \cite{portal1,portal2} the following two concrete models were considered, namely the scalar singlet DM \cite{singlet1,singlet2}, where 
\begin{equation}
\Omega^2(\phi) = 1 + 2 \xi \kappa^2 M \phi
\end{equation}
as well as the inert doublet model \cite{inert1,inert2}, where 
\begin{equation}
\Omega^2(h,\eta) = 1 + 2 \xi \kappa^2 (v + h) \eta
\end{equation}
with $h$ being the SM Higgs boson, $v$ its vacuum expectation value, and $\eta$ the CP-even Higgs boson coming from the second doublet.
Therefore the DM particle may decay into SM particle pairs (if kinematically allowed), $\phi \rightarrow X_1 X_2$, and the precise expressions for the partial decay widths depend on the conformal factor $\Omega^2$ (or on the function $f(\phi)$ if you wish). The lifetime $\tau$ of the DM particle is given by
\begin{equation}
\tau^{-1} = \Gamma_T = \sum_i \Gamma_i
\end{equation}
where the index i runs over all possible decay channels, and $\Gamma_T$ is the total decay width of the scalar field $\phi$. Since the DM particle must be quasi-stable (cosmologically stable), the lifetime of $\phi$ should be higher than the age of the Universe, $t_0 \simeq 4 \times 10^{17}~sec$. In reality, however, the lifetime of the DM is further constrained by telescopes that have been designed to detect its decay products, such as neutrinos (IceCube \cite{icecube}) or $\gamma$ rays (FERMI-Lat telescope \cite{fermi}). In \cite{portal2}, for the models studied there, and for masses $m_\phi \geq 1~GeV$, the authors imposed the conservative limit $\tau \geq 10^{24}~sec$. In the framework we shall be studying here, and if the scalar field $\phi$ is very light, $m_\phi \sim meV$ or lower, as suggested by the cusp/core problem, the only channel kinematically open is the one to photons. However, the scalar field is coupled to massive vector bosons only (see the Feynman rules in the Appendix of \cite{portal2}), and therefore in the scenario adopted here $\phi$ cannot decay into a pair of photons at tree level. The decay process will necessarilly take place at one loop level via charged particles circulating into the loop (this is the case of, for instance, the QCD axion \cite{PQ1,PQ2,axion1,axion2}), and therefore the lifetime of the scalar field is expected to exceed the age of the Universe.

At this point a remark is in order. In the simplified scalar field framework we wish to consider in the present work, $\phi$ does not need to be identified with the DM particle. As a matter of fact, it would be more interesting if we suggested a possible way to constrain the more general case of any new scalar field with certain couplings to electrons. Therefore, in the rest of our work the scalar field does not necessarily serve as the DM particle, keeping the discussion as general as possible. We shall only make a couple of minimal assumptions, namely i) that $\phi$ is a new scalar field beyond the SM of particle physics with a self-interaction potential of the Higgs-like form (\ref{firstkind}), and ii) that it is real and very light, $m_\phi \sim meV$ or lower.

\smallskip

%In the present work we adopt the Inert Doublet model \cite{inert1,inert2,inert3,inert4}, in which there are two Higgs doublets, $H_1,H_2$, and a discrete $Z_2$ symmetry. The second Higgs doublet is inert in the sense that there are no Yukawa couplings with the fermions due to the $Z_2$ symmetry, but they do have gauge interactions as well as self-interactions via the scalar potential. Only the first Higgs doublet acquires a non-vanishing vacuum expectation value (vev), $\langle H_1 \rangle = v$, while the vev of $H_2$ is taken to be zero, $\langle H_2 \rangle = 0$. This ensures the absence of mixing between the components of $H_1$ with those of $H_2$, and therefore $H_1$ closely corresponds to the ordinary SM Higgs doublet. In this case the conformal factor is found to be
%\begin{equation}
%\Omega^2 = 1 + 2 \xi \kappa^2 (v+h) \phi
%\end{equation}
%with $h$ being the SM Higgs boson. In Appendix A.2 of \cite{portal2} the Feynman rules for the interactions vertices may be found. In particular, the Yukawa coupling for the $\phi f \bar{f}$ interaction vertex is found to be
%\begin{equation}
%g = \kappa^2 \xi v m_f
%\end{equation}
%with $m_f$ being the fermion mass. In addition, for a very light scalar field for which the only open decay channels are into massless particles, such as photons,
%the decay rate is computed to be
%\begin{equation}
%\Gamma (\phi \rightarrow \gamma \gamma) = \frac{\xi^2 v^2 \kappa^4 m_{\phi}^3}{32 \pi}
%\end{equation}

\begin{figure}[ht!]
	\centering
	\includegraphics[scale=0.5]{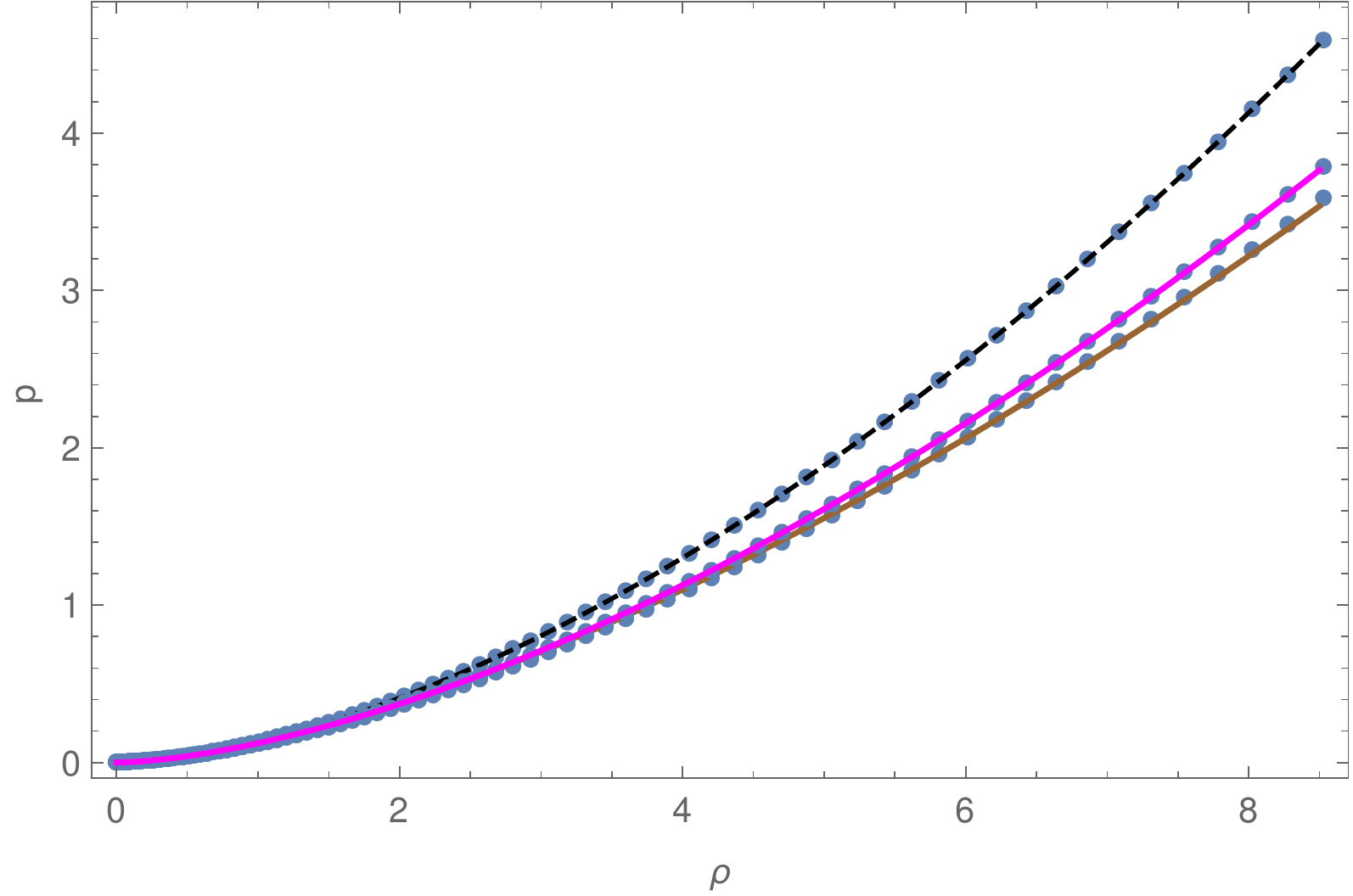}
	\caption{Modification of EoS due to the electron-DM interaction in the gravity portal for $g=4.4 \times 10^{-9}$ and $m_\phi=4.4 \times 10^{-13}~GeV$ (brown curve) and $g=4.1 \times 10^{-9}$ and $m_\phi=4.6 \times 10^{-13}~GeV$ (magenta curve). The standard EoS (black curve) is also shown for comparison reasons.}
	\label{fig:1} 	
\end{figure}

%%%%%%%%%%%%%%%%%%%%%%%%%%%%%%%%%%%%%%%%%%%%%%%%%%%%%%%%%%%%%
\subsection{EoS and cooling time of WDs: Standard treatment}
%%%%%%%%%%%%%%%%%%%%%%%%%%%%%%%%%%%%%%%%%%%%%%%%%%%%%%%%%%%%%

\subsubsection{EoS of an ideal Fermi gas}

White dwarf stars are old compact objects that mark the final evolutionary stage of the vast majority of the stars \cite{ref1,ref2}. Indeed more than $95 \%$, perhaps up to $98 \%$ of all stars, will die as white dwarfs \cite{ref3}. They were discovered in 1914 when H.~Russell noticed that the star now known as 40 Eridani B was located well below the main sequence on the Hertzsprung-Russell diagram. About $80 \%$ of WD show hydrogen atmosphere (DA type), while 20 per cent show helium atmosphere (DB type) \cite{ref4}. The low-mass white dwarfs are expected to harbour He cores, while the average mass white dwarfs most likely contain Carbon/Oxygen cores \cite{ref1}.

At zero-th order approximation, ignoring the Coulomb interactions of electrons, the essential features of the EoS of WDs are captured by the Chandrasekhar model \cite{chandra}. In the standard case without the spin-zero boson, electrons with mass $m$ inside a WD form an ideal Fermi gas, the energy density and pressure of which are given by the well-known expressions \cite{chandra,paper1,paper2}
\begin{eqnarray}
\epsilon_{st} & = & \frac{2}{(2 \pi)^3} \int_0^{k_F} d^3 \vec{k} \sqrt{k^2+m^2} \\
p_{st} & = & \frac{1}{3} \frac{2}{(2 \pi)^3} \int_0^{k_F} d^3 \vec{k} \frac{k^2}{\sqrt{k^2+m^2}}
\end{eqnarray}
where the Fermi wave number $k_F$ is related to the fermion number density $n$ as follows
\be
n = \frac{k_F^3}{3 \pi^2}
\en
The integrals above can be computed exactly, and therefore one can obtain analytical expressions for the pressure and energy density of an ideal Fermi gas as follows
%\begin{widetext}
	\be
	\epsilon_{st}  =  \frac{m^4}{8 \pi^2} \left((x_F+2x_F^3) \sqrt{1+x_F^2}-sinh^{-1}(x_F) \right)
	\en
	\be
	p_{st}  =  \frac{m^4}{24 \pi^2} \left((-3x_F+2x_F^3) \sqrt{1+x_F^2}+3 sinh^{-1}(x_F) \right)
	\en
%\end{widetext} 
where we have defined $x_F=k_F/m$. In addition, we define the scalar baryon density
as follows
\be
n_s = \frac{\partial \epsilon_{st}(m)}{\partial m}=\frac{2}{(2 \pi)^3} \int_0^{k_F} d^3 \vec{k} \frac{m}{\sqrt{k^2+m^2}}
\en
to be useful later on, and it is given by
\be
n_s = \frac{m^3}{2 \pi^2} \left[ x_F \sqrt{1+x_F^2}-ln\left(x_F+\sqrt{1+x_F^2} \right) \right]
\en

In the non-relativistic limit, $x_F \ll 1$, the expression for the pressure takes the approximate form \cite{textbook}
\begin{equation}
p_{st} \sim \frac{m^4 x_F^5}{15 \pi^2}
\end{equation}
while the density is given by \cite{textbook}
\begin{equation}
n = \frac{x_F^3 m^3}{3 \pi^2} = \frac{\rho}{m_u \mu_e}
\end{equation}
where $m_u=1~amu = 1.66 \times 10^{-24}~g$ is the unified atomic mass unit \cite{PDG}, and $\mu_e=A/Z$ , with $A$ being the atomic number of the element of the core, is the molecular weight per electron. There is no explicit dependence on $Z$, and $\mu_e=2$ irrespectively of the core composition \cite{saltas}. Therefore one obtains an EoS of the form
\begin{equation}
p_{st} = \frac{(3 \pi^2)^{5/3}}{15 \pi^2 m (\mu_e m_u)^{5/3}} \: \rho^{5/3} = K_{st} \: \rho^{5/3} 
\end{equation}
where the constant $K_{st}=3.1 \times 10^{12}$ in cgs units.

\begin{figure}[ht!]
	\centering
	\includegraphics[scale=0.6]{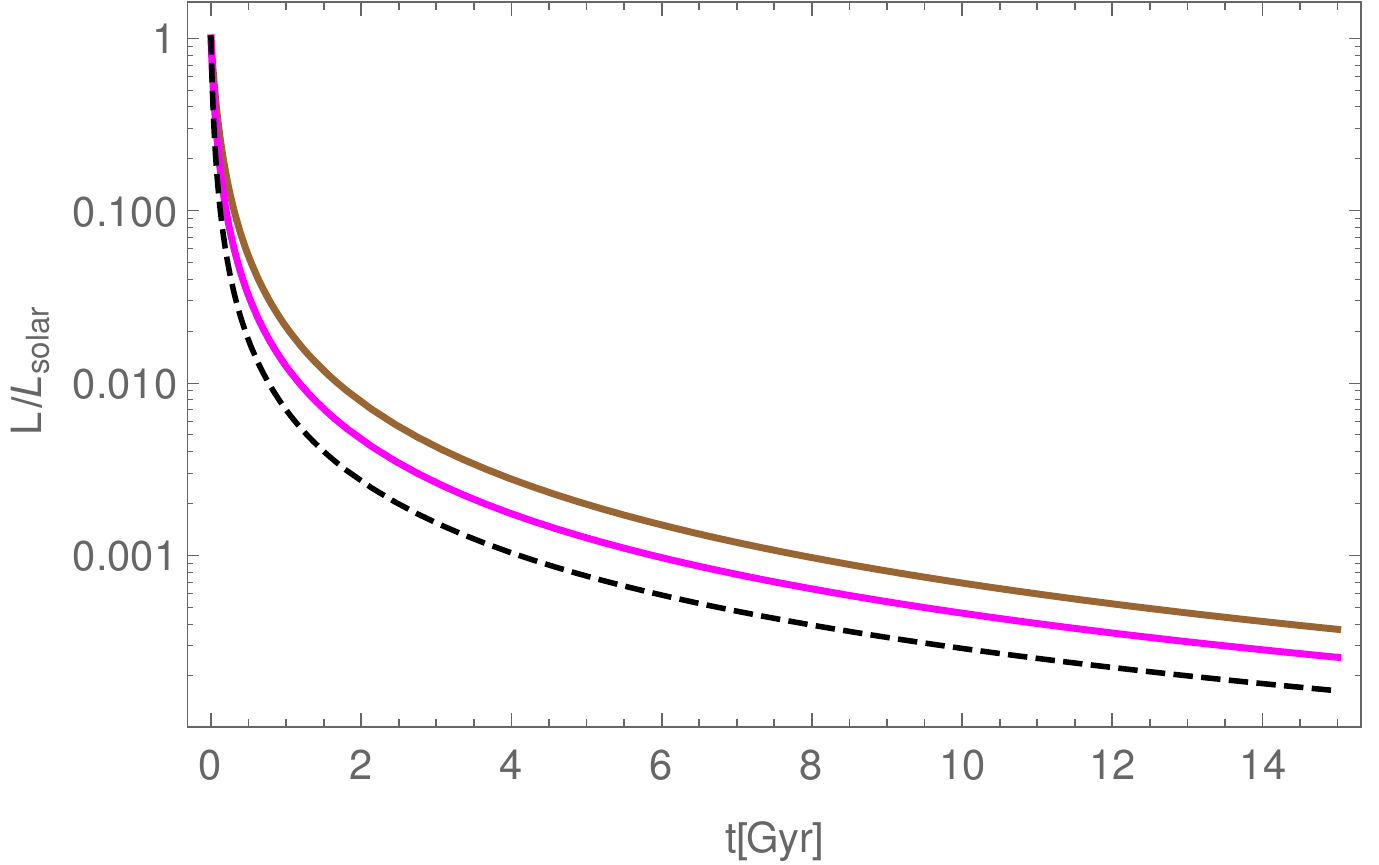}
	\caption{Cooling of white dwarfs in the gravity portal (solid curves) in 
		comparison with its cooling in the standard case (dashed curve) for $M=0.5~M_{\odot}$ and $L_0 = L_{\odot}$.}
	\label{fig:2} 	
\end{figure}

%%%%%%%%%%%%%%%%%%%%%%%%%%%%%%%%%%%%%%%%%%%%%%%%%
\subsubsection{Time dependence of WD luminosity}
%%%%%%%%%%%%%%%%%%%%%%%%%%%%%%%%%%%%%%%%%%%%%%%%%

Since there are no thermonuclear reactions for WD, these objects are cooling down by eradiating, and its emitted energy is due to stored thermal energy. As the pressure drops to zero close to the surface, there must be a non-degenerate atmosphere, which provides an insulating very thin layer (of the order of $10^{-3}$ of the radius of the star) that regulates the rate of heat loss from the object \cite{notes}. The cooling time of WDs in the standard case can be found e.g. in \cite{notes,mestel}. Combining the equations that describe hydrostatic equilibrium and the EoS with the well-known laws \cite{notes}
\begin{eqnarray}
p & = & \frac{ {\cal R} \rho T}{\mu}  \\
L & = & - M C_\nu \frac{dT}{dt}
\end{eqnarray}
where $t$ is the time, $L$ is the luminosity of the WD, $T$ is the constant temperature of its isothermal core, $M$ is the mass of the star, $\mu$ is the mean molecular weight,
 ${\mathcal{R}}=k_B/m_u=8.315 \times 10^7 erg K^{-1} g^{-1}$, with $k_B$ being the Boltzmann constant, is the ideal gas constant \cite{PDG,notes}, and $C_\nu = (3 \mathcal{R})/(2 \mu)$, one finally obtains the time dependence of the WD luminosity
\begin{equation}
\frac{L}{L_0} = \left[1 + \frac{t}{\tau_{st}} \right]^{-7/5}
\label{eq:LLo}
\end{equation}
where $L_0$ is the initial luminosity, while the characteristic cooling time $\tau_{st}$
is computed to be
\begin{equation}
\tau_{st} = \frac{3 {\cal R}}{5 \mu} \left[ \frac{51 \kappa_0 \mu {\cal R}^4}{64 \pi G 4 \sigma \mu_e^5 K_{st}^3} \right]^{2/7} \, \left(\frac{L_0}{M}\right)^{-5/7} 
\label{eq:taust}
\end{equation}
with $M$ being the mass of the WD star, $G$ being Newton's constant, $\kappa_0=2 \times 10^{20}~m^5 kg^{-2} K^{7/2}$ being the opacity \cite{notes}, $\sigma=5.67 \times 10^{-8} W m^{-2} K^{-4}$ being the Stefan-Boltzmann constant \cite{PDG}, and $\mu=1.35$ for realistic compositions \cite{notes}. 
In equation~\ref{eq:LLo} the term $t/\tau_{st}$ is dimensionless.
Accordingly the first and second terms of  equation~\ref{eq:taust}
have units $m^{10/7}\,s^{-8/7}$ and  $m^{-10/7}\,s^{15/7}$,  if the luminosity and mass of the star are express $kg$ and $W$ (in S.I. units).

\begin{figure}[ht!]
	\centering
	\includegraphics[scale=0.6]{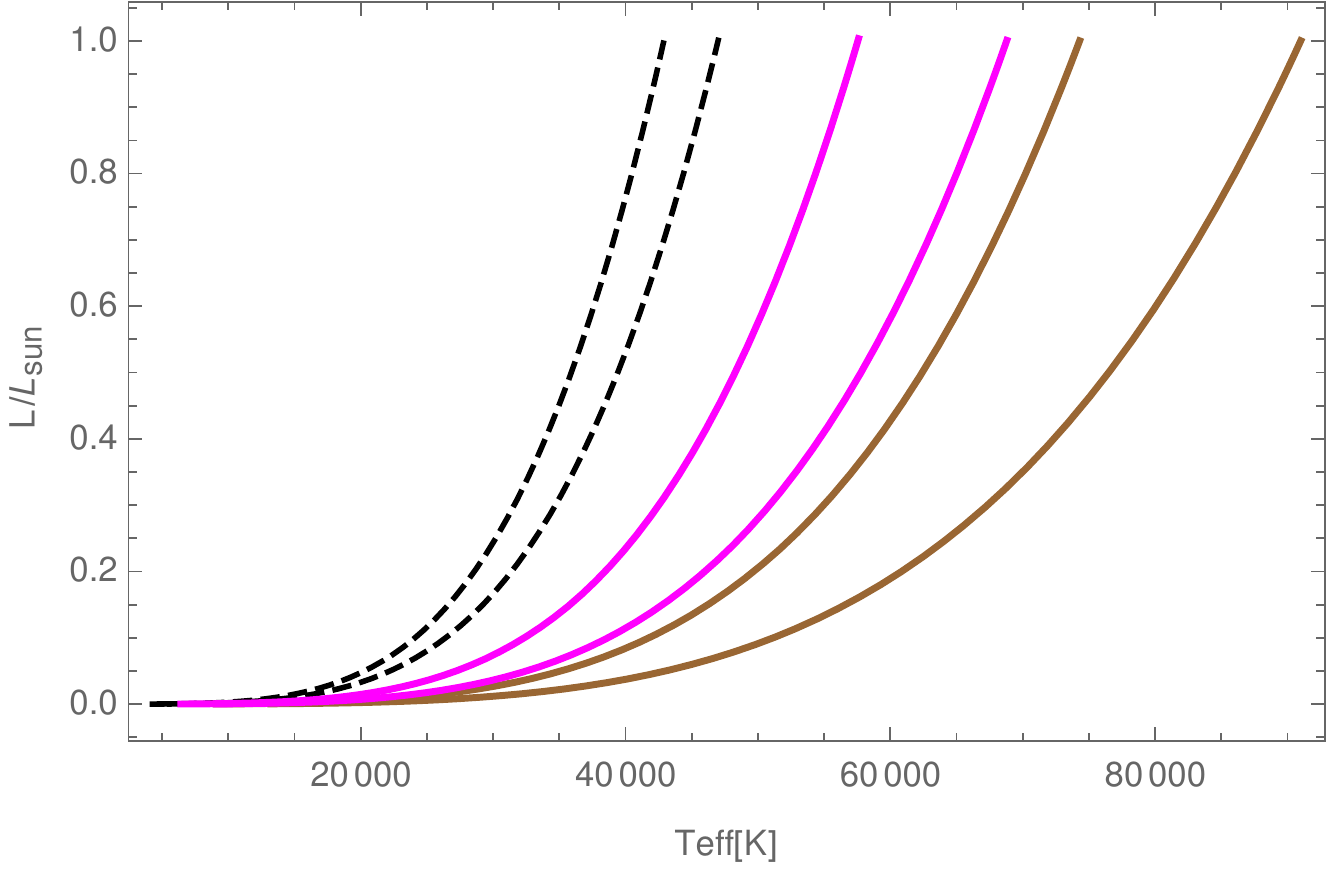}
	\caption{Luminosity of WDs (in units of Solar luminosity) versus effective temperature (in Kelvin) i) for the standard case (in black) and ii) for the two models in the gravity portal scenario (colors as in figures 1 and 2).}
	\label{fig:3} 	
\end{figure}

%%%%%%%%%%%%%%%%%%%%%%%%%%%%%%%%%%%%%%%%%%%%%
\section{Impact of gravity portal on WDs}
%%%%%%%%%%%%%%%%%%%%%%%%%%%%%%%%%%%%%%%%%%%%%

Here we shall study the impact of the DM particle on the EoS of WD and to its cooling time. The treatment is similar to Relativistic Mean Field Theory (RMFT) of neutron stars \cite{walecka1,walecka2,thesis}, where nucleons interact exchanging mesons, the value of which are taken to be constants. In our here we study WD instead of NS, nucleons are replaced by electrons, and finally mesons are replaced by the DM particle $\phi$. 

In the gravity portal, and in the Einstein frame, there is a Yukawa coupling between the DM boson $\phi$ and the electrons, which are Dirac fermions $\psi$. The system is described by the Lagrangian density
%\begin{widetext}
\be
\begin{split}
\mathcal{L}_{e-DM} = \bar{\psi} (i \gamma_\mu \partial^\mu - m + g \phi) \psi + \frac{1}{2} (\partial_\mu \phi \partial^\mu \phi-m_{\phi}^2 \phi^2)
\end{split}
\en
%\end{widetext}
where a possible self-interaction coupling constant $\lambda$ has been ignored, since it will have a negligible effect on the numerical results. The Yukawa coupling constant $g$ depends on the two free parameters of the model, namely the nonminimal coupling $\xi$ and the mass of the scalar field $m_\phi$. It is more convenient, however, to trade $\xi$ for $g$ and take in the following ($m_\phi,g$) to be the two free parameters of the model.

In the mean-field approximation \cite{walecka1,walecka2,thesis} it is assumed that $\phi$ is a constant, $\phi_0$, and therefore the system looks like an ideal Fermi gas where electrons acquire an effective mass
\be
m_* = m - g \phi_0
\en
Since the kinetic term of the DM particle vanishes, the total pressure and energy density of the system is given by
\begin{eqnarray}
p & = & p_{st}(m_*) - \frac{m_{\phi}^2 \phi_0^2}{2} \\
\epsilon & = & \epsilon_{st}(m_*) + \frac{m_{\phi}^2 \phi_0^2}{2}
\end{eqnarray}
where $p_{st}(m_*),\epsilon_{st}(m_*)$ are the standard expressions for the pressure and the energy density respectively of an ideal Fermi gas evaluated at the effective mass $m_*$.
Finally the constant value of the DM boson is given by
\begin{equation}
\phi_0 = \frac{g n_s(m_*)}{m_{\phi}^2}
\end{equation}
where the scalar density $n_s$ is evaluated at the electron effective mass $m_*$. The expression for $\phi_0$ can be obtained from the thermodynamic argument that a closed, isolated system will minimize its energy with respect to the field or the effective mass. 

The effective mass of the electrons is determined solving the equation
\begin{equation}
m_* = m - \frac{g^2 n_s(m_*)}{m_{\phi}^2}
\end{equation}
and the new EoS is obtained.

In Fig.~\ref{fig:1} we show the modification of the EoS in the gravity portal. The points are generated from the numerical solution, while the continuous curves are the fitting curves that correspond to polytropic EoSs $p=K \rho^{(1+1/n)}$ with appropriate $K,n$. In particular, the black curve corresponds to the standard polytropic EoS with index $n=1.5$, 
\begin{equation}
p_{st} = K_{st} \rho^{5/3}
\end{equation}
the brown curve (obtained assuming $g=4.4 \times 10^{-9}$ and $m_\phi=4.4 \times 10^{-13}~GeV$) corresponds to a new EoS with index $n=1.82$,
\begin{equation}
p_1 = K_1 \rho^{1.55} 
\end{equation}
with $K_1=5.75 \times 10^{12}$ in cgs units, and the magenta curve (obtained assuming $g=4.1 \times 10^{-9}$ and $m_\phi=4.6 \times 10^{-13}~GeV$) corresponds to a modified EoS with index $n=1.67$,
\begin{equation}
p_2 = K_2 \rho^{1.6} 
\end{equation}
with $K_2=4.19 \times 10^{12}$ in cgs units.

\smallskip

Going through the same steps for the generic polytropic EoS $p=K \rho^{(1+1/n)}$ one obtains the following
expression for the luminosity
\begin{equation}
\frac{L}{L_0} = \left[1 + \frac{t}{\tau_a} \right]^{1/a}
\end{equation}
where $a$ is found to be
\begin{equation}
a = - \frac{(17/2)-3-2 n}{2 ((17/4)-n-1)},
\end{equation}
while the new characteristic cooling time $\tau_a$ is computed to be
\begin{equation}
\tau_a = \left( \frac{1+a}{-a} \right) \frac{3 {\cal R}}{2 \mu} \left[ \kappa_a \left( \frac{ {\cal R}}{\mu_e} \right)^{2(n+1)} \right]^{1+a} 
\left( \frac{L_0}{M} \right)^{a}
\end{equation}
where  $\kappa_a =(51 \kappa_0 \mu)/(64 \pi G 4 \sigma {\cal R} K^{2n})$.
It is easy to verify that when $n=3/2=1.5$, we recover the expressions of the previous subsection valid in the usual case for $\xi=0=g$. In Fig.~\ref{fig:2} we show the time dependence of the WD luminosity for the standard case and for the modified EoS for $M=0.5~M_{\odot}$ and $L_0 = L_{\odot}$.

Figure~\ref{fig:3} shows the path of WD stars that are going through a cooling process
in a simplified  Hertzsprung-Russell diagram (luminosity decreasing with time).
The two dashed black curves correspond to standard cooling process (as computed using equation \ref{eq:LLo}) of two WDs with masses $0.3 \; M_\odot$ and $0.6 \; M_\odot$. Both stars have initial luminosity $L_0=L_\odot$ as computed using equation \ref{eq:LLo}.
The colour curves correspond to the same stars, but for the models discussed in figures 1 and 2.

It would be interesting to use the predictions of the model and the results obtained here to put constraints on the free parameters of the model using current observational data related to white dwarf stars. We hope to be able to address that issue, and perform a thorough analysis along these lines in a future work.

%%%%%%%%%%%%%%%%%%%%%%%%%%%
\section{Conclusions}
%%%%%%%%%%%%%%%%%%%%%%%%%%%

In summary, in this work we have studied the impact of the gravity portal scenario on the cooling time of white dwarfs. Our results show that the electron-scalar DM interaction leads to a softer equation-of-state, which in turn implies a slower cooling time in comparison with the standard case.

%%%%%%%%%%%%%%%%%%%%%%%%%%%%%%%%%%%%%%%%%%%%%%%%%%%%%%%%%%%%%%%%%%%%%%%%%%

\section*{Acknowlegements}

We wish to thank the anonymous reviewer for useful comments and suggestions.
The authors thank the Funda\c c\~ao para a Ci\^encia e Tecnologia (FCT), Portugal, for the financial support to the Center for Astrophysics and Gravitation-CENTRA, Instituto Superior T\'ecnico, Universidade de Lisboa,  through the Grant No. UID/FIS/00099/2013.

%%%%%%%%%%%%%%%%%%%%%%%%%%%%%%%%%%%%%%%%%%%%%%%%%%%%%%%%%%%%%%%%%%%%%%%%%%

\end{document}